\documentclass{aa}

\usepackage{graphicx}
\usepackage{afterpage}
\usepackage{supertabular}
\usepackage{epsfig}
\begin{document}

\newcommand{\cmr}{~cm$^{-1}$~}
\newcommand{\Teff}{T$_{\rm eff}$~}
\newcommand{\mkm}{$\mu$m~}

\title{Atomic lines in infrared spectra for ultracool dwarfs}

\author{Yuri Lyubchik $^1$, Hugh R.A. Jones$^2$, Yakiv V. Pavlenko$^1$, Serena 
Viti$^{3,4}$, Juliet C. Pickering$^5$, Richard Blackwell-Whitehead$^5$ }

\offprints{Yuri Lyubchik}
\mail{lyu@mao.kiev.ua}

\institute{Main Astronomical Observatory of Academy of Sciences of
Ukraine, Golosiiv woods, Kyiv-127, 03680 Ukraine
\and Astrophysics Research Institute, Liverpool John Moores University, 
Twelve Quays House, Egerton Wharf, Birkenhead CH41 1LD, UK 
\and CNR-Instituto di Fisica dello Spazio Interplanetario, Area di Ricerca di
Tor Vergata, via del fosso del Cavaliere 100, 00133, Roma, Italy
\and Dept of Physics and Astronomy, University College London, Gower St.,
London, WC1E 6BT, UK
\and Blackett Laboratory, Imperial College, Prince Consort Rd, London SW7 2BW, UK}

\date{Received ; accepted }

\authorrunning{Yu.Lyubchik et al.}
\titlerunning{Atomic lines in IR spectra for ultracool dwarfs}


\abstract{We provide a set of atomic lines which are suitable for
the description of ultracool dwarf spectra from 10000 to 25000~\AA.
This atomic linelist was made using both synthetic spectra calculations
and existing atlases of infrared spectra of Arcturus and Sunspot umbra.
We present plots, which show the comparison of synthetic spectra
and observed Arcturus and Sunspot umbral spectra for all atomic lines
likely to be observable in high resolution infrared spectra.
\keywords:{infrared atomic lines -- low-mass objects -- ultracool dwarfs }}

\maketitle

\section{Introduction}

Ultracool dwarfs extend from the cool M dwarfs into the brown dwarf regime
and include the spectral types L and T. It is likely that most of the
ultracool dwarfs are not massive enough to undergo nuclear fusion in their
cores to burn hydrogen and are thus brown dwarfs rather than stars. Here
we are concerned in general terms with the spectroscopy of cool objects
and use the broad temperature classification ``ultracool dwarf'' in 
preference to more
specific terms with a physical meaning such as M dwarf and brown dwarf
(Jones \& Steele 2001).

The astrophysical importance of ultracool dwarf stars derives from the
large-scale cosmological significance of their mass density, to their
extremely slow chemical evolution and to their formation process and their
distinction from planets. Ultracool dwarfs 
span the mass range from the
coolest stars through two orders of magnitude in mass to giant planets.
By determining their fundamental parameters, such as effective
temperature, metallicity and surface gravity, their position in the
Hertzsprung-Russell diagram can be determined. So far these parameters, in
particular surface gravity and metallicity, are poorly
determined. Yet to derive a reliable mass function across the brown dwarf
regime and to discern the abundance patterns of low-metallicity brown
dwarfs, it is essential to design much finer tools of analysis. Changes in
effective temperatures, surface gravity and metallicity can lead to similar
spectral changes. To disentangle these effects requires observations of several
spectral diagnostics that respond differently to changes in these
parameters (e.g., Reid \& Hawley 2000).

Colour information has proved crucial in the identification of most
ultracool dwarfs. 
However, their complex energy distributions mean
that colours do not change monotonically with changing luminosity and
temperature. This makes differentiation of second order effects such as
metallicity and gravity very difficult to discern. Spectroscopic
observations are crucial and low-mass objects have been successfully
classified both in the optical (Kirkpatrick et al. 2000; Basri et al.
2000) and infrared (McLean et al. 2001; 
Geballe et al. 2001; Leggett et al. 2001) regimes across the M and L spectral 
classes. However by the T
spectral class there is insufficient flux at short wavelengths to use
optical wavelengths. Furthermore even by the early L spectral class the
energy distribution at optical wavelengths is almost completely dominated by
broad alkali resonance lines and dust opacities (Pavlenko 1998;
Pavlenko et al. 2000; Burrows et al. 2000; Tsuji 2002). 
The infrared region is also
difficult to model with strong molecular opacities of water, iron hydride
and at cooler temperatures methane as well. However, there are also 
a number of atomic lines observable in the infrared spectra of 
ultracool dwarfs (e.g. McLean et al. 2003). 
These infrared atomic lines lie across the peak in  
energy distribution for these stars  and are thus comparatively
easier to observe than lines of the same atoms in the visible. 

In recent years there has been a focus on improvements to the atomic data
in the optical region, e.g., NIST (Kelleher et al. 1999), Opacity Project 
(Seaton et al. 1992). However, currently
the accuracy and completeness of the atomic data, in particular
oscillator strengths,  in the IR region are far poorer.  In
many cases either the oscillator strengths are highly inaccurate or  there
are simply no measurements available.
This becomes evident when synthetic spectra are generated to match 
infrared spectra, e.g., Jones et al. 1996, Jones \& Viti 2000. 
Here we identify suitable
sensitive infrared atomic lines that may one day be used for reliable 
temperature, 
gravity and abundance analyses for a wide range of ultracool dwarfs, 
from M dwarf stars through brown dwarfs as well as extra-solar 
giant planets. Our paper is the first step in the development of 
a database of infrared atomic lines with reliable oscillator strengths
suitable for the study of low temperature astrophysical objects. Our
specific focus is the interpretation of ultracool dwarfs 
covering the temperature space from the M dwarfs through the brown dwarfs to 
extra-solar giant planets.

Our focus is the prioritisation of atomic lines whose
oscillator strengths we are either already measuring or planning
to measure.
Section 2 describes the observational material, atomic
line lists and theoretical spectra necessary for this study. Our 
procedures for identifying important infrared atomic transitions
are discussed in Section 3. In Section 4 we discuss our results
and conclusions.  In Section 5 we outline our plans for experimental 
measurements of the atomic data. 

\section{Input datasets}

\subsection{Observed spectra of Arcuturus and Sunspot umbra}

For comparisons and the robust identification of atomic lines we used both 
electronic and hard 
copies of the ``Infrared Atlas of the Arcturus spectrum, 0.9 -- 5.3 \mkm'' 
(hereafter ``Arcturus atlas'') (Hinkle et al. 1995) and
``An Atlas of a Dark Sunspot Umbral spectrum from 1970 to 8640 \cmr 
(1.16 to 5.1 \mkm)'' (hereafter ``Sunspot umbra atlas'') (Wallace \& 
Livingston 1992). We used the electronic versions of
both atlases (~ftp://ftp.noao.edu/catalogs/arcturusatlas/ir/ and
ftp://ftp.noao.edu/fts/spot1atl/ ~) to compare with our synthetic spectra.

Arcturus ($\alpha$~Boo) is a K1~III star of effective 
temperature \Teff=4300$\pm$30~K and log~g=1.5$\pm$0.15 (Peterson et al. 
1993). These parameters are far from adequate to describe ultracool dwarf 
atmospheres (~\Teff$\le$2600~K, log~g$\sim$4.5 -- 5.0~)
though the Arcturus atlas provides careful identification of atomic lines and 
gives an opportunity to assess the behaviour of atomic lines when decreasing 
effective temperature from 4000~K to 2000~K.
The atlas contains winter and summer observations. 
Since only the telluric spectrum is affected by seasons, in this paper we 
only use the telluric free summer data. The resolution of this atlas is of the
order of R$\sim$100000.
The presence of a magnetic field in Arcturus is indirectly confirmed by 
strong emission lines in the ultraviolet spectral region formed in 
chromosphere of Arcturus (Ayres et al. 1986).
However the average value of the magnetic 
field is, probably, similar to the solar one, since  
atomic line splitting is not seen in the observed sprectrum.

The Sunspot umbra atlas has a spectral type of M2 -- M5~V, 
which corresponds to an effective temperature range of around 
\Teff=3170 -- 3520~K (Wallace \& Livingston 1992; Viti et al. 1998). 
The magnetic field strength is estimated as $\approx$ 3360 Gauss.
Due to the position of the umbra near the centre of the solar disk 
we see only $\sigma$-components of Zeeman split lines of large 
Lande factor $g$ in the Sunspot umbra atlas.
Due to the strong magnetic field in the spot most of the atomic lines in the 
Sunspot umbral spectrum have doublet-like profiles. Molecular lines, with few 
exceptions, are magnetically 
insensitive and show single profiles. The Sunspot umbra atlas provides a 
spectrum corrected for atmospheric absorption. Its resolution is 
R$\sim$200000 between 10000 and 25000~\AA.

\subsection{Synthetic spectra}

Computations of synthetic spectra were carried out using the WITA6 program
(Pavlenko 2000) assuming Local Thermodynamic Equilibrium. 

For the computation of synthetic specta we used the NextGen 
model atmospheres structures (Hauschildt, private communication, 
Hauschildt et al. 1999) for cool spectra, 
and model atmosphere structures of Kurucz (Kurucz 1994) for the 
higher temperature synthetic spectra of Arcturus.

The atomic linelist used for our spectral modeling and line 
identification was taken from the VALD database (Kupka et al. 1999).
VALD is a compilation of several different lists of atomic
line data which were obtained from experimental measurements and theoretical 
calculations by various authors. 

Chemical equilibrium were computed for the mix of around 100 molecular species. 
Input molecular data and continuum opacity sources
are described elsewhere (Pavlenko et al. 1995; Pavlenko 1997;
Pavlenko 2000). The profiles of absorption lines are described by the 
Voigt function $H(a, v)$. The formulae of Unsold (1955) were used for 
calculation of damping
constants. For these model atmospheres we adopt the conventional
value of microturbulent velocity $v_t$ = 2 km/s. Theoretical spectra were
computed with a wavelength step of 0.05 \AA~ and then convolved with 
gaussians to match the instrumental broadening.

We should note, that we didn't consider dusty effects in our synthetic 
spectra calculations, since the main purpose of the present work 
was not to model
the real spectra of ultracool objects, but to reveal the prominent atomic
features to be observed in such objects. 
The effects of dust will modify atomic line strengths in ultracool dwarf 
spectra. But dust will not affect the quantum-
mechanical parameters which we intend to measure at the present time.

\section{Procedure}

Our procedure of line identification and selection is based on
the comparison of synthetic spectra with the observed spectra of Arcturus and 
Sunspot umbra from  10000 to 25000 \AA. 

For ultracool dwarfs 
we used only synthetic spectra. The choice of model atmosphere
parameters was aimed at obtaining a reliable atomic line list for ultracool 
dwarfs in the IR region. A model atmosphere of 
\Teff/log~g/[M/H] =  2000/5.0/0.0 was selected 
(hereafter ``ultracool dwarf model''). 
This corresponds to an ultracool dwarf 
spectral class of around L0 (Kirkpatrick et al. 1999, Martin et al. 1999). The 
half width of the gaussian we used to account for
instrumental broadening in the ``ultracool dwarf'' 
synthetic spectrum is 0.4\AA~ 
which corresponds to resolution 
R$\sim$20000 -- 40000, approximately the resolution available to modern
infrared echelle spectrometers operating between 10000 -- 25000\AA. 
Another model atmosphere 4000/4.5/0.0 (hereafter 
``Arcturus-like model'') was chosen to select some observable spectral atomic 
features formed in hotter atmospheres. The half width of the gaussian for
this synthetic spectrum is 0.13\AA~ to match the line broadening by 
macroturbulent velocity 
$v_{\rm macro}$=3.5 km/s in the Arcturus spectrum (Peterson et al. 1993).
The \Teff = 4000~K is slightly lower than Arcturus 
\Teff=4300~K (Peterson et al. 1993), and higher than the temperature of 
the Sunspot umbra \Teff$\sim$3170 -- 3520~K (Wallace \& Livingston 1992).

In the electronic versions of the Arcturus and Sunspot umbra atlases the data 
are given on a wavenumber scale. We transform the wavenumbers to wavelengths 
in air using the formula from Allen's Astrophysical Quantaties (2000).

It is well known that the difficulties of line identification in IR spectra
of ultracool dwarfs have a number of causes.
There are many strong molecular lines of H$_2$O (Jones et al. 2002), FeH 
(Cushing et al. 2003), CO (Pavlenko \& Jones 2003), CH$_4$ (Noll et al. 2000)
native to ultracool dwarfs as well as telluric
molecular lines, particulary strong between the J, H and K bands. 
Nevertheless, we were able to identify many absorption features in the
atmospheres of Arcturus and Sunspot umbra spectra likely to be 
present in ultracool dwarfs.
To do this, we developed a procedure based on the inter-comparison 
of computed and observed spectra designed to minimise possible errors in 
line identification. 
The line identification files were made using \\
a) ``ultracool dwarf''  synthetic spectrum (2000/5.0/0.0), \\
b) ``Arcturus-like''  synthetic spectrum (4000/4.5/0.0).\\
In addition we used line identification files from hard copy Arcturus and 
Sunspot umbra atlases.

In our identification procedure we assume that lines absorb only
at a given wavelength. This assumption provides an upper limit of 
absorption. In 
the real spectra the residual intensity may be lower due to contribution of 
other lines.
For every line from the VALD linelist in the range  of 10000 --
25000\AA~ the residual flux $r_{\nu}= 
\frac{({\rm Flux~in~line})_{\nu}}{({\rm Flux~in~continuum})_{\nu}}$
was computed. 
Atomic lines from VALD with 
central intensities deeper than 0.8 of the residual flux in the spectrum 
of 2000/5.0/0.0 atmosphere
were selected to restrict the number of lines to those likely to be observable
in cool spectra.
These lines are given in the Table~1. The set of the selected lines is 
rather large 
and thus we assign different levels of priority to the lines.
The main criteria for selecting lines of the highest priority is that the 
central intensity of the identified line ought to be greater than 0.6 of 
the residual flux (column 4 in the Table~1) in the ``ultracool dwarf'' 
synthetic spectrum.

Plots for all spectral regions containing atomic lines of interest are shown in 
Figure~1 ( they are also available at 
http:// www.astro.livjm.ac.uk/ $^{\sim}$hraj/ spectralatlas/ index.html ). The 
identifications of atomic features are labelled 
in the spectra. Arrows and labels at the top of the 
plots show identification in the ``ultracool dwarf'' 
synthetic spectrum, the same labels
in the bottom of the plots are for the "Arcturus-like" spectrum. 
The Sunspot umbral spectrum is shown by bold line, the 
Arcturus spectrum by a dotted line, the ``ultracool dwarf'' 
synthetic spectrum as a solid line and ``Arcturus-like'' 
synthetic spectrum as a dashed line. In order to simplify our plots molecular 
line identifications are not included. Usually 
molecular lines have narrow deep profiles allowing them to be distinguished 
from atomic lines. Their identification in the Sunspot umbra and Arcturus 
can be found in the hard copy atlases.

\section{Results}

Our results are shown in two tables.
Table~1 contains all identified lines with central intensities 
deeper than 0.8 of residual flux using ``ultracool dwarf model''
identifications. Table~2 investigates the sensitivity of priority 1 lines
identified in Table~1 to tempereture, gravity and metallicity.

In the first column of Table~1 we estimate the priority of 
lines for measurement of atomic data. \\ 
a) Priority 1 indicates that atomic line is deeper than 0.6 
in the 2000/5.0/0.0 synthetic spectrum. Also we include as first priority 
three lines of Ti (22211.229, 22232.838, 22274.012\AA) which are
shallower than 0.6. They are located far from strong atomic and 
molecular lines and thus are good for identification purposes in the 
spectra of ultracool dwarfs. \\ 
b) Priority 2 indicates that the residual flux of an atomic line is between 
0.8 -- 0.6.\\ 
c) Priority 3 indicates that the atomic line is located in the wing of a
nearby stronger line or in a wavelength region with strong telluric 
absorption. 

Atomic lines without an assigned priority might be relevant in the 
analysis of ultracool dwarf 
spectra, but they are predicted to be weak in the synthetic spectrum. 

The second column shows the wavelength of the line in \AA~ in air . These 
values were taken from VALD. We check that wavelengths from Arcturus and 
Sunspot umbra 
atlases identification files are in a good agreement with VALD.
The wavelengths of a few lines differ from VALD values by less 
than 0.1\AA~ in the region 10000 -- 20000 \AA~ and by less than 
0.3\AA~ beyond 20000 \AA.  

The third column shows the atomic identifications of the line.

Columns 4 -- 7 show the linedepths of identified lines. Column 4 -- 
the residual flux in the line obtained from the ``ultracool dwarf'' 
synthetic spectrum; 
column 5 -- from the "Arcturus-like" synthetic spectrum; column 6 -- measured 
from the hard copy of Arcturus 
atlas depth of identified atomic lines; column 7 -- the same as column 6 for 
the hard copy of the Sunspot umbra atlas. 
As was mentioned above we used the hard copy versions 
of the Arcturus and Sunspot umbra atlases to measure linedepths. 
The accuracy of our line depth measurements from the hardcopy atlases is 
$\sim$0.01 -- 0.02 for the Arcturus atlas (column 6) and $\sim$0.02 -- 0.03 
for the Sunspot umbra atlas (column 7). 
We note that linedepth information given in columns 6 and 7 
is only used to check the presence of a line in the observed spectra.
Thus we do not need  high accuracy linedepth measurements for the  
observable spectra of the Sunspot umbra and Arcturus. 

We note that the data in the Sunspot umbra atlas provide a spectrum from 
11600\AA, rather than 10000\AA~ where our figures and tables start from. 
Some lines are labelled (see column 7) by the symbol "$\sim$" 
which means that part of the line in the hard copy of the atlas is obscured by 
a gap in the data and the value given in the table is for the visible part of 
the line. In columns 6 and 7 there are also some lines with "bl" suffices. 
This means that these lines are blended with stronger nearby lines.

There are two regions with strong telluric absorption ( marked in columns 6 and 
7 by $\bigoplus$): 13600--14770\AA~ and 18260--19450\AA~, regions between 
the J and H, and H and K photometric bands. Lines 
identified from VALD in these regions are difficult to
observe in ground based observations but can be used in probable 
future space observations of ultracool dwarfs. 

Finally we note, that in Table~1 we italicise atomic lines of 
Rb, Y, Ba and Lu, identified only 
using theoretical computations. Although these lines are not seen in the 
observations of Arcturus and Sunspot umbra, they are strong in synthetic 
spectrum computed for the 2000/5.0/0.0 model atmosphere, 
and therefore also likely to be strong below 2000~K.

In Table~2 we show only lines which were marked in Table~1 as first 
priority lines. 
The first column are wavelengths in \AA; the second column are names of elements.
In the third column we give the central line intensity for the ``ultracool 
dwarf'' (2000/5.0/0.0) spectrum.
We compute some synthetic spectra for ultracool objects varying parameters 
of \Teff, log~g and metallicity to show the sensitivity of the central 
intensities of identified lines to these parameters. 
For our computations we chose the minimal step in model atmospheres grid:
$\Delta$\Teff = 100~K, $\Delta$log~g = 0.5, $\Delta$[M/H]=-0.5.
In columns 4 -- 6 we 
show the difference of central intensity ( in \% ) computations with ``new''
models from central intensity computations for our ``ultracool dwarf model''. 
Negative values in columns 4 -- 6 indicate that a line is stronger in the 
new model relative to the reference model 2000/5.0/0.0.

Column 4 shows the dependence of residual fluxes on effective temperature. 
For comparison 
a 2100/5.0/0.0 synthetic spectrum was used. One can see that only a few 
lines show strong temperature dependence for $\Delta$\Teff=100~K. 
The NextGen 2000/4.5/0.0 model was used with 2000/5.0/0.0 to test gravity 
sensitivity. The dependence for $\Delta$(log~g)=0.5 is also relatively 
weak for most lines and is shown in column 5. 
$\Delta$[M/H]=0.5 from computations using NextGen model atmospheres 
2000/5.0/-0.5 and 2000/5.0/0.0 is shown in column 6.
The metallicity dependence is relatively high for all lines. 
Thus Table~2 indicates that relatively small variations of model 
atmosphere parameters will
not seriously affect our line selection and prioritisation.

\section{Experimental work}

 The identification and prioritisation of atomic lines  presented in this 
paper is only the first step in our project and is primarily based on a
theoretical treatment. 
Our program of measurements of atomic data for the priority lines given in 
this paper is underway.  We are currently measuring oscillator strengths 
by measurements of sets of relative line intensities using the high resolution 
Fourier
transform spectrometers at Imperial College (Pickering 2002) and NIST (National 
Institute of Standards and
Technology, US).  
We have already recorded spectra of Ti, Mn and Na, and measurements of Mg, K, 
Ca and  Fe are in progress.  Measurements of other species listed in Table 1 
are underway in the near future. 
These line intensities will be used to obtain branching ratios which are then
combined with level lifetimes to obtain f-values. We expect to achieve 
uncertainties in f-values of around 10-15 \%.  

\section{Acknowledgements}

We would like to thank the authors of the Arcturus and Sunspot umbra atlases 
for making 
their data available through an ftp site and Prof. Hauschildt and the 
PHOENIX team
for NextGen model atmospheres. NSO/Kitt Peak FTS data used here 
were produced 
by NSF/NOAO. SV thanks the Italian Space Agency (ISA) for financial support.
RBW thanks Gillian Nave at NIST for the assistance and facilities provided for 
the IR experimental measurements.
All the authors thank the Royal Society and PPARC of the UK for 
support of travel and experimental work.

\newpage
\begin{table}
\caption[]{Atomic lines found in ultracool dwarf synthetic spectrum from 
10000 -- 25000\AA.
The columns are priority of the line (Pr), wavelength in \AA ($\lambda$ (\AA)),
element name (Atom),
residual flux in ``ultracool dwarf'' synthetic spectrum (UCD\_ss),
residual flux in ``Arcturus-like'' synthetic spectrum (Ar\_ss),
residual line intensity from hard copy of Arcturus atlas (Ar\_At),
residual line intensity from hard copy of Sunspot umbra atlas (Ss\_At).}
\end{table}
\footnotesize
\tablefirsthead{\hline 
 Pr & $\lambda$ (\AA) & Atom & UCD\_ss & Ar\_ss & Ar\_At & Ss\_At \\ \hline } 
\tablehead{\hline  
 Pr & $\lambda$ (\AA) & Atom & UCD\_ss & Ar\_ss & Ar\_At & Ss\_At \\ \hline }  
\tabletail{\hline }
\tablelasttail{\hline }
\begin{supertabular}{|c|c|c|c|c|c|c|}
    & 10003.086 & Ti~I & 0.909  & 0.716  & 0.75 &    --  \\
    & 10011.745 & Ti~I & 0.928  & 0.756  & 0.77 &    --  \\
    &{\it 10032.099} &{\it Ba~I} &{\it 0.887} &{\it 1.000} &{\it 0.97} &  --  \\
    & 10034.491 & Ti~I & 0.855  & 0.739  & 0.63 &    --  \\
    & 10048.826 & Ti~I & 0.845  & 0.730  & 0.65 &    --  \\
    & 10057.728 & Ti~I & 0.867  & 0.659  & 0.65 &    --  \\
    & 10059.904 & Ti~I & 0.868  & 0.779  & 0.72 &    --  \\
    & 10155.159 & Fe~I & 0.917  & 0.790  & 0.67 &    --  \\
    & 10167.463 & Fe~I & 0.893  & 0.753  & 0.65 &    --  \\
2   & 10340.880 & Fe~I & 0.772  & 0.608  & 0.55 &    --  \\
1   & 10343.820 & Ca~I & 0.365  & 0.420  & 0.44 &    --  \\
    & 10378.995 & Fe~I & 0.904  & 0.747  & 0.68 &    --  \\
3   & 10395.789 & Fe~I & 0.650  & 0.566  & 0.50 &    --  \\
1   & 10396.802 & Ti~I & 0.384  & 0.441  & 0.35 &    --  \\
    &{\it 10471.272} &{\it Ba~I} &{\it 0.936}  &{\it 1.000} &{\it 0.99} & --  \\
1   & 10496.114 & Ti~I & 0.403  & 0.445  & 0.35 &    --  \\
1   & 10584.633 & Ti~I & 0.425  & 0.449  & 0.38 &    --  \\
2   & 10607.718 & Ti~I & 0.677  & 0.639  & 0.70 &    --  \\
1   & 10661.623 & Ti~I & 0.455  & 0.475  & 0.41 &    --  \\
2   & 10677.047 & Ti~I & 0.621  & 0.581  & 0.61 &    --  \\
1   & 10726.391 & Ti~I & 0.489  & 0.490  & 0.45 &    --  \\
2   & 10732.864 & Ti~I & 0.615  & 0.591  & 0.60 &    --  \\
2   & 10746.440 & Na~I & 0.783  & 0.840  & 0.94 &    --  \\
    & 10749.293 & Na~I & 0.880  & 0.749  & 1.00 &    --  \\
2   & 10774.866 & Ti~I & 0.652  & 0.622  & 0.63 &    --  \\
2   & 10821.176 & S~I  & 0.696  & 0.451  & 0.85 &    --  \\
    & 10827.896 & Ti~I & 0.928  & 0.924  & 0.97 &    --  \\
2   & 10834.848 & Na~I & 0.757  & 0.732  & 0.84 &    --  \\
    & 10847.634 & Ti~I & 0.926  & 0.926  & 0.95 &    --  \\
1   & 11019.848 & K~I  & 0.477  & 0.902  & 1.00 &    --  \\
1   & 11022.653 & K~I  & 0.550  & 0.934  & 0.95 &    --  \\
2   & 11119.798 & Fe~I & 0.764  & 0.536  & 0.52 &    --  \\ 
2   & 11251.116 & Fe~I & 0.766  & 0.546  & 0.98 &    --  \\
    & 11298.862 & Fe~I & 0.837  & 0.609  & 0.55 &    --  \\
    &{\it 11303.052} &{\it Ba~I} &{\it 0.488} &{\it 0.977} &{\it 1.00} & --  \\
3   & 11374.081 & Fe~I & 0.557  & 0.538  & 0.50 &    --  \\
1   & 11381.454 & Na~I & 0.094  & 0.416  & 0.35 &    --  \\
1   & 11403.779 & Na~I & 0.076  & 0.385  & 0.26 &    --  \\
3   & 11422.323 & Fe~I & 0.509  & 0.469  & 0.37 &    --  \\
2   & 11439.127 & Fe~I & 0.650  & 0.476  &  --  &    --  \\ 
1   & 11593.591 & Fe~I & 0.477  & 0.437  &  --  &   0.70 \\
1   & 11607.575 & Fe~I & 0.359  & 0.389  & 0.30 &   0.60 \\ 
    & 11610.522 & Cr~I & 0.899  & 0.557  & 0.52 &   0.70 \\
1   & 11638.264 & Fe~I & 0.395  & 0.411  & 0.33 & $\sim$0.55 \\
1   & 11689.976 & Fe~I & 0.021  & 0.388  & 0.31 &   0.45 \\
1   & 11690.220 & K~I  & 0.013  & 0.483  & 0.40bl & $\sim$0.48 \\
1   & 11769.639 & K~I  & 0.061  & 0.601  & 0.70 &   0.65 \\ 
1   & 11772.838 & K~I  & 0.004  & 0.468  & 0.42 &  $\sim$0.55 \\
3   & 11780.543 & Ti~I & 0.522  & 0.736  & 0.75 &   0.82 \\ 
3   & 11783.267 & Fe~I & 0.485  & 0.453  & 0.40 &   0.67 \\
3   & 11797.179 & Ti~I & 0.779  & 0.757  & 0.75 &   0.82 \\
2   & 11828.185 & Mg~I & 0.705  & 0.430  & 0.21 &   0.50 \\
1   & 11882.847 & Fe~I & 0.277  & 0.353  & 0.29 &   0.55 \\
1   & 11884.085 & Fe~I & 0.364  & 0.398  & 0.33 &   0.57 \\
2   & 11892.878 & Ti~I & 0.731  & 0.606  & 0.57 &   0.70 \\
2   & 11949.542 & Ti~I & 0.700  & 0.571  & 0.55 &   0.72 \\
    & 11955.955 & Ca~I & 0.878  & 0.620  & 0.85 &   0.82 \\
1   & 11973.050 & Fe~I & 0.225  & 0.334  & 0.28 &    --  \\
1   & 11973.848 & Ti~I & 0.498  & 0.574  & 0.53 &  $\sim$0.50 \\
1   & 12432.273 & K~I  & 0.080  & 0.582  & 0.67 &   0.65 \\
2   & 12484.618 & Ti~I & 0.793  & 0.661  & 0.98 &   0.98 \\
1   & 12522.134 & K~I  & 0.022  & 0.528  & 0.57 &   0.60 \\ 
3   & 12526.081 & K~I  & 0.429  & 0.989  & 0.97 &   0.99 \\
2   & 12556.999 & Fe~I & 0.790  & 0.606  & 0.62 &   0.80 \\
    & 12600.277 & Ti~I & 0.862  & 0.748  & 0.75 &   0.80 \\
    & 12671.092 & Ti~I & 0.864  & 0.746  & 0.75 &   0.83 \\
2   & 12679.144 & Na~I & 0.656  & 0.589  & 0.66 &   0.65 \\
2   & 12679.224 & Na~I & 0.658  & 0.604  & 0.66 &   0.65 \\
    & 12738.383 & Ti~I & 0.914  & 0.726  & 0.75 &   0.75 \\
    & 12816.046 & Ca~I & 0.857  & 0.615  & 0.79 &   0.76 \\
2   & 12821.672 & Ti~I & 0.628  & 0.519  & 0.45 &   0.67 \\
    & 12823.868 & Ca~I & 0.889  & 0.671  & 0.83 &   0.78 \\
2   & 12831.442 & Ti~I & 0.645  & 0.527  & 0.50 &   0.70 \\
2   & 12847.033 & Ti~I & 0.621  & 0.512  & 0.50 &   0.70 \\
2   & 12879.769 & Fe~I & 0.740  & 0.571  & 0.55 &   0.75 \\
1   & 12899.764 & Mn~I & 0.468  & 0.444  & 0.22 &   0.57 \\
    & 12975.912 & Mn~I & 0.851  & 0.587  & 0.46 &   0.75 \\
    & 13011.895 & Ti~I & 0.852  & 0.729  & 0.75 &   0.77 \\
    & 13077.263 & Ti~I & 0.896  & 0.798  & 0.70 &   0.80 \\
1   & 13123.410 & Al~I & 0.433  & 0.390  & 0.21 &   0.45 \\
1   & 13150.753 & Al~I & 0.500  & 0.415  & 0.25 &   0.52 \\
    &{\it 13233.279} &{\it Rb~I} &{\it 0.642} &{\it 1.000} &{\it 1.00} &{\it 0.99} \\
    & 13281.485 & Mn~I & 0.902  & 0.632  & 0.55 &   0.82 \\
2   & 13293.795 & Mn~I & 0.619  & 0.497  & 0.33 &   0.72 \\
1   & 13318.944 & Mn~I & 0.547  & 0.465  & 0.27 &   0.67 \\
    &{\it 13371.782} &{\it Lu~I} &{\it 0.385} &{\it 0.749} &{\it 1.00} & --  \\
    & 13377.794 & K~I  & 0.854  & 0.989  & 1.00 &    --  \\
    & 13409.108 & Ti~I & 0.716  & 0.583  & 0.97 &   1.00 \\
    & 13600.372 & Mn~I & 0.818  & 0.627  & $\bigoplus$  & $\bigoplus$  \\
3   & 13626.696 & Mn~I & 0.593  & 0.477  & $\bigoplus$  & $\bigoplus$  \\
3   & 13642.930 & Mn~I & 0.672  & 0.519  & $\bigoplus$  & $\bigoplus$  \\
    &{\it 13662.993} &{\it Rb~I} &{\it 0.523} &{\it 0.989} & $\bigoplus$  & $\bigoplus$ \\
    & 13688.911 & Ti~I & 0.905  & 0.822  & $\bigoplus$  & $\bigoplus$  \\
    & 13787.736 & V~I  & 0.832  & 0.746  & $\bigoplus$  & $\bigoplus$  \\
3   & 13847.454 & Mn~I & 0.725  & 0.558  & $\bigoplus$  & $\bigoplus$  \\
3   & 13864.219 & Mn~I & 0.625  & 0.492  & $\bigoplus$  & $\bigoplus$  \\
3   & 13997.520 & Mn~I & 0.663  & 0.509  & $\bigoplus$  & $\bigoplus$  \\
    & 14040.849 & V~I  & 0.841  & 0.757  & $\bigoplus$  & $\bigoplus$  \\
    & 14253.141 & V~I  & 0.848  & 0.765  & $\bigoplus$  & $\bigoplus$  \\
    & 14270.253 & Ca~I & 0.850  & 0.531  & $\bigoplus$  & $\bigoplus$  \\
    & 14308.700 & Fe~I & 0.852  & 0.551  & $\bigoplus$  & $\bigoplus$  \\       
    & 14423.053 & V~I  & 0.875  & 0.807  & $\bigoplus$  & $\bigoplus$  \\
    &{\it 14752.870} &{\it Rb~I} &{\it 0.368} &{\it 0.968} & $\bigoplus$ & $\bigoplus$  \\
2   & 14767.484 & Na~I & 0.786  & 0.706  & $\bigoplus$  & $\bigoplus$  \\
2   & 14779.732 & Na~I & 0.706  & 0.615  & 0.99 &   0.57bl \\
    &{\it 14999.917} &{\it Ba~I} &{\it 0.435} &{\it 0.952} &{\it 1.00} &{\it 0.95} \\
2   & 15024.992 & Mg~I & 0.780  & 0.399  & 0.24 &   0.49 \\ 
    & 15040.246 & Mg~I & 0.815  & 0.412  & 0.24 &   0.55 \\
    & 15057.099 & Ca~I & 0.829  & 0.895  & 0.97 &   0.77 \\
2   & 15077.291 & Fe~I & 0.771  & 0.545  & 0.98 &   0.79 \\
1   & 15163.067 & K~I  & 0.254  & 0.534  & 0.67 &   0.67 \\
1   & 15163.067 & K~I  & 0.254  & 0.534  & 0.67 &   0.67 \\
1   & 15168.376 & K~I  & 0.280  & 0.572  & 0.70 &   0.72 \\
    &{\it 15288.938} &{\it Rb~I} &{\it 0.501} &{\it 0.993} &{\it 1.00} &{\it 0.85} \\
    &{\it 15289.966} &{\it Rb~I} &{\it 0.299} &{\it 0.940} &{\it 1.00} &{\it 0.90} \\
2   & 15334.847 & Ti~I & 0.618  & 0.436  & 0.45 &   0.60 \\
    & 15490.339 & Fe~I & 0.858  & 0.642  & 0.64 &   1.00 \\
2   & 15543.756 & Ti~I & 0.628  & 0.451  & 0.47 &   0.65 \\
    & 15602.842 & Ti~I & 0.867  & 0.646  & 0.73 &   0.72 \\
    & 15698.979 & Ti~I & 0.855  & 0.699  & 0.70 &   0.75 \\
2   & 15715.573 & Ti~I & 0.629  & 0.446  & 0.50 &   0.67 \\
    & 15836.788 & Ti~I & 0.851  & 0.710  & 0.73 &   0.85 \\
2   & 16136.823 & Ca~I & 0.795  & 0.474  & 0.76 &   0.70 \\
2   & 16150.763 & Ca~I & 0.695  & 0.427  & 0.68 &   0.67 \\
    & 16155.236 & Ca~I & 0.829  & 0.496  & 0.80 &   0.80 \\
2   & 16157.364 & Ca~I & 0.674  & 0.416  & 0.67 &   0.62 \\
2   & 16197.075 & Ca~I & 0.635  & 0.398  & 0.55 &   0.56 \\
    & 16204.087 & Ca~I & 0.825  & 0.494  & 0.85 &    --  \\
2   & 16388.857 & Na~I & 0.771  & 0.770  & 0.95 &   0.87 \\
    & 16454.533 & Si~I & 0.868  & 0.585  & 0.98 &   0.93 \\
2   & 16718.957 & Al~I & 0.678  & 0.398  & 0.55 &   0.45bl \\
2   & 16750.564 & Al~I & 0.626  & 0.380  & 0.27 &   0.50 \\
2   & 17108.662 & Mg~I & 0.799  & 0.403  & 0.27 &   0.47 \\
    &{\it 17123.809} &{\it Y~I} &{\it 0.581} &{\it 0.699} &{\it 0.99} &{\it 0.96} \\
    &{\it 17422.836} &{\it Y~I} &{\it 0.537} &{\it 0.631} &{\it 1.00} &{\it 0.90} \\
    &{\it 17663.289} &{\it Y~I} &{\it 0.526} &{\it 0.625} &{\it 0.95} &{\it 0.98} \\
    &{\it 17903.229} &{\it Y~I} &{\it 0.489} &{\it 0.559} &{\it 1.00} &  --  \\
    &{\it 18049.807} &{\it Y~I} &{\it 0.473} &{\it 0.557} &{\it 1.00} &  --  \\
    &{\it 18115.295} &{\it Y~I} &{\it 0.458} &{\it 0.533} &{\it 1.00} &  --  \\
    &{\it 18181.758} &{\it Y~I} &{\it 0.450} &{\it 0.539} &  --  &    --  \\
    &{\it 18234.975} &{\it Lu~I} &{\it 0.377} &{\it 0.732} &{\it 1.00} &  --  \\
    &{\it 18293.682} &{\it Y~I}  &{\it 0.523} &{\it 0.641} & $\bigoplus$  & $\bigoplus$ \\
1   & 18465.312 & Na~I & 0.333  & 0.402  & $\bigoplus$  & $\bigoplus$  \\
1   & 18465.312 & Na~I & 0.333  & 0.402  & $\bigoplus$  & $\bigoplus$  \\
1   & 18465.484 & Na~I & 0.337  & 0.410  & $\bigoplus$  & $\bigoplus$  \\
3   & 18856.646 & Fe~I & 0.770  & 0.458  & $\bigoplus$  & $\bigoplus$  \\
    & 18987.010 & Fe~I & 0.830  & 0.495  & $\bigoplus$  & $\bigoplus$  \\
    & 19046.178 & Ca~I & 0.828  & 0.555  & $\bigoplus$  & $\bigoplus$  \\
1   & 19309.223 & Ca~I & 0.144  & 0.326  & $\bigoplus$  & $\bigoplus$  \\
1   & 19452.982 & Ca~I & 0.125  & 0.311  & $\sim$0.20 &    -- \\
1   & 19505.738 & Ca~I & 0.142  & 0.337  & 0.22 &  $\sim$0.30 \\
    &{\it 19728.117} &{\it Y~I} &{\it 0.601} &{\it 0.726} &{\it 0.99} &{\it 0.95} \\
1   & 19776.771 & Ca~I & 0.074  & 0.306  & 0.16 &   0.30 \\
1   & 19853.092 & Ca~I & 0.516  & 0.428  & 0.45 &   0.50 \\
1   & 19862.191 & Ca~I & 0.162  & 0.345  & 0.25 &  $\sim$0.35 \\
1   & 19917.195 & Ca~I & 0.280  & 0.417  & 0.45 &   0.65 \\
1   & 19933.729 & Ca~I & 0.555  & 0.443  & 0.45 &   0.65 \\
2   & 19961.832 & Ca~I & 0.638  & 0.477  & $\sim$0.60 &    -- \\
    &{\it 20170.207} &{\it Y~I} &{\it 0.647} &{\it 0.778} &{\it 1.00} &{\it 1.00} \\
    &{\it 20831.520} &{\it Y~I} &{\it 0.624} &{\it 0.747} &{\it 1.00} &{\it 0.95} \\
    &{\it 20879.256} &{\it Y~I} &{\it 0.678} &{\it 0.812} &{\it 0.97} &{\it 0.90} \\
2   & 21163.756 & Al~I & 0.794  & 0.497  & 0.45 &   0.65 \\
    &{\it 21260.449} &{\it Y~I} &{\it 0.526} &{\it 0.617} &{\it 0.96} &{\it 1.00} \\
    &{\it 21574.170} &{\it Y~I} &{\it 0.605} &{\it 0.733} &{\it 1.00} &{\it 0.95} \\
    & 21730.506 & Sc~I & 0.821  & 0.904  & 0.95 &   0.98 \\
1   & 21782.926 & Ti~I & 0.532  & 0.467  & 0.50 &   0.70 \\
    & 21812.055 & Sc~I & 0.802  & 0.896  & 0.95 &   0.93 \\
2   & 21897.377 & Ti~I & 0.583  & 0.493  & 0.55 &   0.72 \\
2   & 22004.494 & Ti~I & 0.671  & 0.551  & 0.66 &   0.80 \\
1   & 22052.141 & Sc~I & 0.493  & 0.579  & 0.75 &   0.65 \\
1   & 22056.426 & Na~I & 0.342  & 0.469  & 0.50 &   0.50 \\
1   & 22065.232 & Sc~I & 0.544  & 0.635  & 0.83 &   0.75 \\
1   & 22083.662 & Na~I & 0.383  & 0.487  & 0.56 &   0.60 \\
1   & 22211.229 & Ti~I & 0.650  & 0.537  & 0.63 &   0.80 \\
1   & 22232.838 & Ti~I & 0.628  & 0.521  & 0.61 &   0.76 \\
2   & 22266.729 & Sc~I & 0.648  & 0.756  & 0.90 &   0.85 \\
1   & 22274.012 & Ti~I & 0.654  & 0.540  & 0.64 &   0.75 \\
3   & 22310.617 & Ti~I & 0.682  & 0.613  & 0.75 &   0.80 \\
    &{\it 22311.588} &{\it Ba~I} &{\it 0.475} &{\it 0.980} &{\it 0.98} &{\it 0.96} \\
2   & 22394.670 & Sc~I & 0.587  & 0.686  & 0.89 &   0.81 \\ 
2   & 22443.904 & Ti~I & 0.773  & 0.657  & 0.77 &   0.85 \\
    &{\it 22543.826} &{\it Y~I} &{\it 0.538} &{\it 0.647} &{\it 1.00} &{\it 0.97} \\
    & 22656.963 & Sc~I & 0.844  & 0.919  & 0.97 &   0.90 \\
2   & 22963.336 & Ti~I & 0.675  & 0.547  & 0.65 &   0.75 \\
2   & 22986.250 & Sc~I & 0.605  & 0.697  & 0.89 &   0.87 \\
    &{\it 23204.070} &{\it Y~I} &{\it 0.686} &{\it 0.826} &{\it 1.00} &{\it 0.96} \\ 
    &{\it 23253.688} &{\it Ba~I} &{\it 0.467} &{\it 0.966} &{\it 0.98} &{\it 0.98} \\ 
1   & 23348.424 & Na~I & 0.563  & 0.548  & 0.60bl & 0.64 \\
1   & 23378.945 & Na~I & 0.516  & 0.535  &  --  &   0.60 \\
1   & 23379.137 & Na~I & 0.513  & 0.525  & $\sim$0.60 &  0.60 \\
2   & 23404.758 & Sc~I & 0.646  & 0.755  & 0.90  &  0.95 \\
2   & 23441.471 & Ti~I & 0.702  & 0.572  & 0.60bl & 0.70 \\
2   & 23719.029 & Ti~I & 0.736  & 0.595  &  --  &   0.98 \\
    & 23809.379 & Sc~I & 0.694  & 0.801  & 0.96 &   0.95 \\
    &{\it 23990.445} &{\it Y~I} &{\it 0.554} &{\it 0.675} &{\it 1.00} &  --  \\ 
    & 24074.787 & Sc~I & 0.734  & 0.839  & 0.93 &   0.87 \\
    &{\it 24170.594} &{\it Lu~I} &{\it 0.190} &{\it 0.630} &{\it 0.97} &{\it 0.95} \\
2   & 24281.809 & Ti~I & 0.788  & 0.611  & 1.00 &   0.70 \\ 
\end{supertabular}
\vspace{5mm}

\newpage
\begin{table}
\caption[]{ Sensitivity of first priority lines to temperature, gravity and 
metallicity.
The columns are wavelength in \AA ($\lambda$ (\AA)), name of element (Atom), 
central line intensity in the 2000/5.0/0.0 synthetic spectrum (LD),
temperature sensitivity in \% ($\Delta r_{\nu}^{\rm (Teff)}$),
log~g sensitivity in \% ($\Delta r_{\nu}^{\rm (log~g)}$), 
metallicity sensitivity in \% ($\Delta r_{\nu}^{\rm ([M/H])}$). }
\end{table}
\tablefirsthead{\hline 
$\lambda$ (\AA) & Atom & LD &$\Delta r_{\nu}^{\rm (Teff)}$ & $\Delta r_{\nu}^{\rm (log~g)}$ & $\Delta r_{\nu}^{\rm ([M/H])}$\\ 
\hline }  
\tablehead{\hline 
$\lambda$ (\AA) & Atom & LD &$\Delta r_{\nu}^{\rm (Teff)}$ & $\Delta r_{\nu}^{\rm (log~g)}$ & $\Delta r_{\nu}^{\rm ([M/H])}$\\ 
\hline }
\tabletail{\hline }
\tablelasttail{\hline }
\begin{supertabular}{|c|c|c|c|c|c|} 
 10343.820 & Ca~I & 0.172 & -1.22 &  -6.86 &   19.49 \\ 
 10396.802 & Ti~I & 0.115 &  4.27 &  -2.01 &  -19.01 \\ 
 10496.114 & Ti~I & 0.125 &  4.24 &  -2.08 &  -18.65 \\ 
 10584.633 & Ti~I & 0.138 &  4.06 &  -2.18 &  -18.21 \\ 
 10661.623 & Ti~I & 0.155 &  3.56 &  -2.33 &  -17.85 \\ 
 10726.391 & Ti~I & 0.176 &  3.02 &  -2.50 &  -17.36 \\ 
 11019.848 & K~I  & 0.423 & -4.75 & -26.12 &   55.45 \\ 
 11022.653 & K~I  & 0.489 & -4.11 & -25.90 &   48.95 \\ 
 11381.454 & Na~I & 0.039 &  5.94 &  -8.27 &   27.39 \\ 
 11403.779 & Na~I & 0.030 &  6.98 &  -7.64 &   23.92 \\ 
 11593.591 & Fe~I & 0.186 &  3.70 &  -3.38 &   25.00 \\ 
 11607.575 & Fe~I & 0.119 &  5.42 &  -3.17 &   22.44 \\ 
 11638.264 & Fe~I & 0.136 &  5.22 &  -3.23 &   23.51 \\ 
 11689.976 & Fe~I & 0.135 &  5.12 &  -3.04 &   23.01 \\ 
 11690.220 & K~I  & 0.013 & 14.39 &  -7.58 &   18.94 \\ 
 11769.639 & K~I  & 0.027 & 11.44 &  -7.75 &   22.51 \\ 
 11772.838 & K~I  & 0.011 & 15.74 &  -6.48 &   19.44 \\ 
 11882.847 & Fe~I & 0.094 & 6.08  &  -2.99 &   19.55 \\ 
 11884.085 & Fe~I & 0.139 & 4.96  &  -3.02 &   22.66 \\ 
 11973.050 & Fe~I & 0.078 & 6.43  &  -3.21 &   17.74 \\ 
 11973.848 & Fe~I & 0.462 & 0.02  &  -3.98 &   -2.38 \\ 
 12432.273 & K~I  & 0.032 &  9.18 & -10.13 &   29.43 \\ 
 12522.134 & K~I  & 0.025 & 10.08 &  -9.68 &   25.00 \\ 
 12899.764 & Mn~I & 0.219 &  3.10 &  -4.05 &   24.90 \\ 
 13123.410 & Al~I & 0.319 & -4.76 &  -5.51 &   16.89 \\ 
 13150.753 & Al~I & 0.370 & -2.73 &  -6.24 &   19.58 \\ 
 13318.944 & Mn~I & 0.299 &  2.17 &  -4.71 &   26.03 \\ 
 15163.067 & K~I  & 0.449 & -1.32 & -12.69 &   39.75 \\ 
 15163.067 & K~I  & 0.188 & -0.05 &  -7.47 &   29.71 \\ 
 15168.376 & K~I  & 0.208 & -0.58 &  -8.22 &   31.19 \\ 
 18465.312 & Na~I & 0.575 & -2.66 & -11.25 &   22.02 \\ 
 18465.312 & Na~I & 0.353 & -2.69 &  -7.12 &   20.91 \\ 
 18465.484 & Na~I & 0.376 & -2.68 &  -8.02 &   20.94 \\ 
 19309.223 & Ca~I & 0.091 &  2.76 &   ---  &    4.85 \\ 
 19452.982 & Ca~I & 0.078 &  3.08 &  -7.82 &    3.85 \\ 
 19505.738 & Ca~I & 0.100 &  2.60 &  -7.69 &    5.39 \\ 
 19776.771 & Ca~I & 0.073 &  3.03 &  -7.71 &    3.44 \\ 
 19853.092 & Ca~I & 0.506 & -3.16 &  -5.64 &   13.11 \\ 
 19862.191 & Ca~I & 0.107 &  2.54 &  -7.51 &    5.54 \\ 
 19917.195 & Ca~I & 0.196 &  1.43 &  -7.56 &   12.57 \\ 
 19933.729 & Ca~I & 0.537 & -1.81 &  -6.34 &   14.41 \\ 
 21782.926 & Ti~I & 0.479 & -1.46 &  -4.14 &    0.17 \\ 
 22052.141 & Sc~I & 0.374 &  2.22 &  -8.95 &   29.38 \\ 
 22056.426 & Na~I & 0.311 & -0.77 &  -5.17 &   17.47 \\ 
 22065.232 & Sc~I & 0.424 &  1.91 &  -9.68 &   30.26 \\ 
 22083.662 & Na~I & 0.356 & -2.50 &  -5.42 &   17.85 \\ 
 22211.229 & Ti~I & 0.589 & -0.93 &  -6.59 &    6.50 \\
 22232.838 & Ti~I & 0.569 & -0.88 &  -6.01 &    5.41 \\
 22274.012 & Ti~I & 0.594 & -0.91 &  -6.65 &    6.64 \\
 23348.424 & Na~I & 0.564 & -2.73 &  -7.47 &   16.79 \\ 
 23379.137 & Na~I & 0.526 & -3.71 &  -7.71 &   16.45 \\ 
\end{supertabular}

\end{document}